\documentclass[11pt]{article}
\usepackage{latexsym,amssymb,amsmath}
\begin{document}
\baselineskip=18pt

\newcommand{\la}{\langle}
\newcommand{\ra}{\rangle}
\newcommand{\psp}{\vspace{0.4cm}}
\newcommand{\pse}{\vspace{0.2cm}}
\newcommand{\ptl}{\partial}
\newcommand{\dlt}{\delta}
\newcommand{\sgm}{\sigma}
\newcommand{\al}{\alpha}
\newcommand{\be}{\beta}
\newcommand{\G}{\Gamma}
\newcommand{\gm}{\gamma}
\newcommand{\vs}{\varsigma}
\newcommand{\Lmd}{\Lambda}
\newcommand{\lmd}{\lambda}
\newcommand{\td}{\tilde}
\newcommand{\vf}{\varphi}
\newcommand{\yt}{Y^{\nu}}
\newcommand{\wt}{\mbox{wt}\:}
\newcommand{\rd}{\mbox{Res}}
\newcommand{\ad}{\mbox{ad}}
\newcommand{\stl}{\stackrel}
\newcommand{\ol}{\overline}
\newcommand{\ul}{\underline}
\newcommand{\es}{\epsilon}
\newcommand{\dmd}{\diamond}
\newcommand{\clt}{\clubsuit}
\newcommand{\vt}{\vartheta}
\newcommand{\ves}{\varepsilon}
\newcommand{\dg}{\dagger}
\newcommand{\tr}{\mbox{Tr}}
\newcommand{\ga}{{\cal G}({\cal A})}
\newcommand{\hga}{\hat{\cal G}({\cal A})}
\newcommand{\Edo}{\mbox{End}\:}
\newcommand{\for}{\mbox{for}}
\newcommand{\kn}{\mbox{ker}}
\newcommand{\Dlt}{\Delta}
\newcommand{\rad}{\mbox{Rad}}
\newcommand{\rta}{\rightarrow}
\newcommand{\mbb}{\mathbb}
\newcommand{\lra}{\Longrightarrow}

\begin{center}{\Large \bf Multiple Parameter Function Approaches to the }\end{center}
\begin{center}{\Large \bf
Equations of Dynamic Convection in a Sea}

\end{center}

\begin{center}{\large Xiaoping Xu}\end{center}
\begin{center}{Institute of Mathematics, Academy of Mathematics \& System Sciences}\end{center}
\begin{center}{Chinese Academy of Sciences, Beijing 100190, P.R. China}
\footnote{Research supported
 by China NSF 10871193}\end{center}

\vspace{0.3cm}

 \begin{center}{\Large\bf Abstract}\end{center}

\vspace{0.2cm} {\small One of the most important topics in
geophysics is to study convection in a sea. Based on the algebraic
characteristics of the equations of dynamic convection in a sea, we
introduce various schemes with multiple parameter functions to solve
these equations and obtain families of new explicit exact solutions
with multiple parameter functions. Moreover, symmetry
transformations are used to simplify our arguments.}

\section{Introduction}

Both the atmospheric and oceanic flows are  influenced by the
rotation of the earth. In fact, the fast rotation and small aspect
ratio are two main characteristics of the large scale atmospheric
and oceanic flows. The small aspect ratio characteristic leads to
the primitive equations, and the fast rotation leads to the
quasi-geostropic equations (cf. [2], [7], [8], [10]). A main
objective in climate dynamics and in geophysical fluid dynamics is
to understand and predict the periodic, quasi-periodic, aperiodic,
and fully turbulent characteristics of the large scale atmospheric
and oceanic flows (e.g., cf. [4], [6]).

The general model of atmospheric and oceanic flows is very
complicated. Various simplified models had been established and
studied. For instance, Boussinesq equations are simpler models in
atmospheric sciences (e.g., cf. [9]). Chae [1] proved the global
regularity, and Hou and Li [3] obtained the well-posedness of the
two-dimensional equations.  Hsia, Ma and Wang [4] studied the
bifurcation and periodic solutions of the three-dimensional
equations.

 The following  equations in geophysics
$$u_x+v_y+w_z=0,\qquad \rho=p_z,\eqno(1.1)$$
$$\rho_t+u\rho_x+v\rho_y+w\rho_z=0,\eqno(1.2)$$
$$u_t+uu_x+vu_y+wu_z+v=-\frac{1}{\rho}p_x,\eqno(1.3)$$
$$v_t+uv_x+vv_y+wv_z-u=-\frac{1}{\rho}p_y,\eqno(1.4)$$
are used to describe the dynamic convection in a sea, where $u,\:v$
and $w$ are components of velocity vector of relative motion of
fluid in Cartesian coordinates $(x,y,z)$, $\rho=\rho(x,y,z,t)$ is
the density of fluid and $p$ is the pressure (e.g., cf. Page 203 in
[5]). Ovsiannikov  determined the Lie point symmetries of the above
equations and found two very special solutions (cf. [5]).

In [11], we used the stable range of nonlinear term to solve the
equation of nonstationary transonic gas flow. Moreover, we [12]
solved the three-dimensional Navior-Stokes equations by asymmetric
techniques and moving frames. Based on the algebraic characteristics
of the equations (1.1)-(1.4) of dynamic convection in a sea, we
introduce various schemes with multiple parameter functions to solve
these equations and obtain families of new explicit exact solutions
with multiple parameter functions. Moreover, symmetry
transformations are used to simplify our arguments. By specifying
these parameter functions, one can obtain the solutions of certain
practical models.

For convenience, we always assume that all the involved partial
derivatives of related functions always exist and we can change
orders of taking partial derivatives. We also use prime $'$ to
denote the derivative of any one-variable function. We will use the
following symmetry transformations $T_1$-$T_4$ due to Ovsiannikov (
cf. Page 204 of [5]) of the equation (1.1)-(1.4) to simplify our
solutions:
$$T_1(u)=u(t,x+\al,y,z+{\al'}'x-\al'y)-\al',\qquad T_1(v)=
v(t,x+\al,y,z+{\al'}'x-\al'y),\eqno(1.5)$$
$$T_1(w)=w(t,x+\al,y,z+{\al'}'x-\al' y)-{\al'}'u+\al'v-{{\al'}'}'x+{\al'}'y,\qquad
\eqno(1.6)$$
$$T_2(u)=u(t,x,y+\al,z+\al'x+{\al'}'y),\qquad T_2(v)=
v(t,x,y+\al,z+\al' x+{\al'}'y)-\al',\eqno(1.7)$$
$$T_2(w)=w(t,x,y+\al,z+\al'x+{\al'}'y)-\al'u-{\al'}'v-{\al'}'x-{{\al'}'}'y,\qquad
\eqno(1.8)$$
$$T_1(p)=p(t,x+\al,y,z+{\al'}'x-\al'y),\qquad
T_2(p)=p(t,x,y+\al,z+\al'x+{\al'}'y),\eqno(1.9)$$
$$T_3(u)=u(t,x,y,z+\al),\qquad T_3(v)=v(t,x,y,z+\al),\eqno(1.10)$$
$$T_3(w)=w(t,x,y,z+\al)-\al',\qquad T_3(p)=p(t,x,y,z+\al),\eqno(1.11)$$
$$T_4(p)=p+\al,\qquad T_4(F)=F\qquad\for\;\;F=u,v,w,\eqno(1.12)$$
where $\al$ is an arbitrary function of $t$. The above
transformations transform one solution of the equations (1.1)-(1.4)
into another solution. Applying the above transformations to any
solution found in this paper will yield another solution with four
extra parameter functions.

In Section 2, we use a new variable of moving line to solve the
equations (1.1)-(1.4). An approach of using the product of
cylindrical invariant function with $z$ is introduced in Section 3.
In Section 4, we reduce the three-dimensional (spacial) equations
(1.1)-(1.4) into a two-dimensional problem and then solve it with
three different ansatzes.

\section{Moving-Line Approach}

Let $\al$ and $\be$ be given functions of $t$. Denote
$$\varpi=\al'x+\be'y+z.\eqno(2.1)$$
Suppose that $f,g,h$ are functions in $t,x,y,z$ that are linear in
$x,y,z$ such that
$$f_x+g_y+h_z=0.\eqno(2.2)$$
 We assume
$$u=\phi(t,\varpi)+f,\qquad v=\psi(t,\varpi)+g,\eqno(2.3)$$
$$w=h-\al'\phi(t,\varpi)- \be'\psi(t,\varpi),\qquad
p=\zeta(t,\varpi),\eqno(2.4)$$ where $\phi,\psi,\zeta$ are
two-variable functions. Note that the first equation in (1.1)
naturally holds and  $\rho=p_z=\zeta_\varpi$ by the second
equation in (1.1). Moreover, (1.2)-(1.4) become
$$\zeta_{\varpi
t}+\zeta_{\varpi\varpi}({\al'}'x+{\be'}'y+\al'f+\be'g+h)=0,\eqno(2.5)$$
\begin{eqnarray*}\hspace{2cm}& &f_t+g+ff_x+gf_y+hf_z+\al'+\phi_t
+(f_x-\al'f_z)\phi +(f_y-\be'f_z+1)\psi\\ &
&+\phi_\varpi({\al'}'x+{\be'}'y+\al'f+\be'g+h)
=0,\hspace{6cm}(2.6)\end{eqnarray*}
\begin{eqnarray*}\hspace{1cm}& &g_t-f+fg_x+gg_y+hg_z+\be'+\psi_t
+(g_x-\al'g_z-1)\phi +(g_y-\be'g_z)\psi\\ &
&+\psi_\varpi({\al'}'x+{\be'}'y+\al'f+\be'g+h)
=0.\hspace{7cm}(2.7)\end{eqnarray*}

In order to solve the above system of partial differential
equations, we assume
$${\al'}'x+{\be'}'y+\al'f+\be'g+h=-\gm'\varpi=-\gm'(\al'x+\be'y+z)
\eqno(2.8)$$ for some function $\gm$ of $t$, and
$$f_t+g+ff_x+gf_y+hf_z+\al'=0,\eqno(2.9)$$
$$g_t-f+fg_x+gg_y+hg_z+\be'=0.\eqno(2.10)$$
Then (2.5)-(2.7) become
$$\zeta_{\varpi
t}-\gm'\varpi\zeta_{\varpi\varpi}=0,\eqno(2.11)$$
$$\phi_t
+(f_x-\al'f_z)\phi
+(f_y-\be'f_z+1)\psi-\gm'\varpi\phi_\varpi=0,\eqno(2.12)$$
$$\psi_t+(g_x-\al'g_z-1)\phi
+(g_y-\be'g_z)\psi-\gm'\varpi\psi_\varpi=0.\eqno(2.13)$$

According to (2.8),
$$h=-{\al'}'x-{\be'}'y-\al'f-\be'g-\gm'\varpi.\eqno(2.14)$$
Substituting the above equation into (2.9) and (2.10), we have:
$$f_t+f(f_x-\al'f_z)+g(f_y-\be'f_z+1)-f_z({\al'}'x+{\be'}'y+\gm'\varpi)
+\al'=0,\eqno(2.15)$$
$$g_t+f(g_x-\al'g_z-1)+g(g_y-\be'g_z)-g_z({\al'}'x+{\be'}'y+\gm'\varpi)
+\be'=0.\eqno(2.16)$$ Our linearity assumption implies that
$$A=\left(\begin{array}{cc}f_x-\al'f_z&f_y-\be'f_z+1\\
g_x-\al'g_z-1&g_y-\be'g_z\end{array}\right)\eqno(2.17)$$ is a
matrix function of $t$. In order to solve the system (2.12) and
(2.13), and the system (2.15) and (2.16), we need the
commutativity of $A$ with $dA/dt$. For simplicity, we assume
$$f_y-\be'f_z+1=g_x-\al'g_z-1=0.\eqno(2.18)$$
So
$$f_y=\be'f_z-1,\qquad g_x=\al'g_z+1.\eqno(2.19)$$
Moreover, (2.15) and (2.16) become
$$f_t+f(f_x-\al'f_z)-f_z({\al'}'x+{\be'}'y+\gm'\varpi)
+\al'=0,\eqno(2.20)$$
$$g_t+g(g_y-\be'g_z)-g_z({\al'}'x+{\be'}'y+\gm'\varpi)
+\be'=0.\eqno(2.21)$$ Write
$$f=\al_1x+(\be'\al_2-1)y+\al_2 z+\al_3,\eqno(2.22)$$
$$g=(\al'\be_2+1)x+\be_1y+\be_2 z+\be_3\eqno(2.23)$$
by our linearity assumption and (2.19), where $\al_i$ and $\be_j$
are functions of $t$.

 Now (2.20) is equivalent to the following
system of ordinary differential equations:
$$\al_1'+\al_1(\al_1-\al'\al_2)-\al_2({\al'}'+\gm'\al')=0,\eqno(2.24)$$
$$(\be'\al_2)'+(\be'\al_2-1)(\al_1-\al'\al_2)
-\al_2({\be'}'+\gm'\be')=0,\eqno(2.25)$$
$$\al_2'+\al_2(\al_1-\al'\al_2-\gm')=0,\eqno(2.26)$$
$$\al_3'+\al_3(\al_1-\al'\al_2)+\al'=0.\eqno(2.27)$$
Observe that $(2.25)-\be'\times(2.26)$ becomes
$$-\al_1+\al'\al_2=0.\eqno(2.28)$$
So (2.26) becomes
$$\al_2'-\gm'\al_2=0\lra \al_2=b_1e^\gm,\qquad
b_1\in\mbb{R}.\eqno(2.29)$$ According to (2.28),
$$\al_1=b_1\al'e^\gm.\eqno(2.30)$$
With the data (2.29) and (2.30), (2.24) naturally holds. By
(2.27), we take
$$\al_3=-\al.\eqno(2.31)$$

Note that (2.21) is equivalent to the following system of ordinary
differential equations:
$$\al'\be_2'+(\al'\be_2+1)(\be_1-\be'\be_2)-
\al'\be_2\gm'=0,\eqno(2.32)$$
$$\be_1'+\be_1(\be_1-\be'\be_2)-\be_2({\be'}'+\be'\gm')=0,\eqno(2.33)$$
$$\be_2'+\be_2(\be_1-\be'\be_2-\gm')=0,\eqno(2.34)$$
$$\be_3'+\be_3(\be_1-\be'\be_2)-\be'=0.\eqno(2.35)$$
Similarly, we have:
$$\be_1=b_2\be'e^\gm,\qquad\be_2=b_2e^\gm,\qquad
\be_3=\be\eqno(2.36)$$ with $b_2\in\mbb{R}$. Moreover, (2.2) gives
$\gm'=0$ by (2.14), (2.28) and (2.36). We take $\gm=0$. Therefore,
$\phi=\Im(\varpi)$ and $\psi=\iota(\varpi)$ by (2.12) and (2.13)
for some one-variable functions $\Im$ and $\iota$. Furthermore, we
take $\zeta=\sgm(\varpi)$ by (2.11) for another one-variable
function $\sgm$. In summary, we have:\psp

{\bf Theorem 2.1}. {\it Let $\al,\be$ be functions of $t$ and let
$b_1,b_2\in\mbb{R}$. Suppose that $\Im,\;\iota$ and $\sgm$ are
arbitrary one-variable functions. The following is a solution of
the equations (1.1)-(1.4) of  dynamic convection in a sea:
$$u=b_1\al'x+(b_1\be'-1)y+b_1
z-\al+\Im(\al'x+\be'y+z),\eqno(2.37)$$
$$v=(b_2\al'+1)x+b_2\be' y+b_2 z+\be+
\iota(\al'x+\be'y+z),\eqno(2.38)$$
\begin{eqnarray*}w&=&-({\al'}'+b_1(\al')^2+(b_2\al'+1)\be')x
-({\be'}'+\al'(b_1\be'-1)+b_2(\be')^2)y-(b_1\al'+b_2\be')z\\
&
&+\al\al'-\be\be'-\al'\Im(\al'x+\be'y+z)-\be'\iota(\al'x+\be'y+z),
\hspace{4.5cm}(2.39)\end{eqnarray*}
$$p=\sgm(\al'x+\be'y+z),\qquad\rho=\sgm'(\al'x+\be'y+z).\eqno(2.40)$$
}\pse

We remark that we have tried some other forms of the matrix $A$ in
(2.17) such that $A$ and $dA/dt$ commute, but we have failed to
get new solutions.

\section{Approach of Cylindrical Product}

Let $\sgm$ be a fixed one-variable function and set
$$\varpi=\sgm(x^2+y^2)z.\eqno(3.1)$$
Suppose that $f$ and $g$  are functions in $t,x,z$ that are linear
homogeneous in $x,y$ and
$$h=\frac{\gm}{\sgm}-z(f_x+g_y),
\eqno(3.2)$$ where $\gm$ is a function of $t$. Assume
$$u=f+y\psi(t,\varpi),\qquad v=g-x\psi(t,\varpi),\qquad
w=h,\qquad p=\phi(t,\varpi)\eqno(3.3)$$ where $\psi$ and $\phi$
are two-variable functions. Note
$$u_t=f_t+y\psi_t,\qquad u_x=f_x+2xyz\sgm'\psi_\varpi,\eqno(3.4)$$
$$u_y=f_y+\psi+2y^2z\sgm'\psi_\varpi,\qquad
u_z=f_z+y\sgm\psi_\varpi,\eqno(3.5)$$
$$v_t=g_t-x\psi_t,\qquad v_x=g_x-\psi-2x^2z\sgm'
\psi_\varpi,\eqno(3.6)$$
$$v_y=g_y-2xyz\sgm'\psi_\varpi,\qquad
v_z=g_z-x\sgm\psi_\varpi.\eqno(3.7)$$ Hence (1.3) becomes
\begin{eqnarray*}\hspace{1.3cm}&&u_t+uu_x+vu_y+wu_z+v=f_t+y\psi_t+(f+y\psi)
(f_x+2xyz\sgm'\psi_\varpi)\\ &
&+(g-x\psi)(f_y+1+\psi+2y^2z\sgm'\psi_\varpi)+y\sgm h\psi_\varpi
\\ &=&f_t+ff_x+g(1+f_y)+x(g_x-f_y-1)\psi-x\psi^2\\ & &+y[\psi_t
+(f_x+g_y)\psi+(2(xf+yg)\sgm'z+h\sgm)
\psi_\varpi]=-\frac{2xz\sgm'}{\sgm}\hspace{3.2cm}(3.8)
\end{eqnarray*}
and (1.4) gives
\begin{eqnarray*}\hspace{1.3cm}&&v_t+uv_x+vv_y+wv_z-u=g_t-x\psi_t+(f+y\psi)
(g_x-1-\psi-2x^2z\sgm'\psi_\varpi)\\ &
&+(g-x\psi)(g_y-2xyz\sgm'\psi_\varpi)-x\sgm h\psi_\varpi
\\ &=&g_t+f(g_x-1)+gg_y-y(1+f_y-g_x)\psi-y\psi^2\\ & &-x[\psi_t+(f_x+g_y)\psi
+(2(xf+yg)\sgm'z+h\sgm)
\psi_\varpi]=-\frac{2yz\sgm'}{\sgm}.\hspace{3.1cm}(3.9)
\end{eqnarray*}

In order to solve the above system of differential equations, we
assume
$$f=\al' x-\frac{y}{2},\qquad g= \frac{x}{2}+\al' y,\qquad
\sgm(x^2+y^2)=\frac{1}{x^2+y^2}\eqno(3.10)$$ for some function
$\al$ of $t$. According to (3.2),
$$h=\frac{\gm}{\sgm}-2\al' z.
\eqno(3.11)$$ Now (3.8) becomes
$$({\al'}'+(\al')^2+4^{-1}-\psi^2)x+y[\psi_t+2\al'\psi+(\gm-4\al'\varpi)
\psi_\varpi]=2x\varpi\eqno(3.12)$$ and (3.9) yields
$$({\al'}'+(\al')^2+4^{-1}-\psi^2)y-x[\psi_t+2\al'\psi+(\gm-4\al'\varpi)
\psi_\varpi]=2y\varpi.\eqno(3.13)$$ The above system is equivalent
to
$${\al'}'+(\al')^2+4^{-1}-\psi^2=2\varpi,\eqno(3.14)$$
$$\psi_t+2\al'\psi+(\gm-4\al'\varpi)\psi_\varpi=0.\eqno(3.15)$$
By (3.14), we take
$$\psi=\sqrt{{\al'}'+(\al')^2+4^{-1}-2\varpi},\eqno(3.16)$$
due to the skew-symmetry of $(u,x)$ and $(v,y)$. Substituting
(3.16) into (3.15), we get
$${{\al'}'}'+2\al'{\al'}'+4\al'({\al'}'+(\al')^2+4^{-1}-2\varpi)
-2(\gm-4\al'\varpi)=0, \eqno(3.17)$$ equivalently,
$$\gm=2(\al')^3+3\al'{\al'}'+\frac{{{\al'}'}'+\al'}{2}.\eqno(3.18)$$

According to the second equation in (1.1), we have
$\rho=\sgm\phi_\varpi$. Note
$$\rho_t=\sgm\phi_{\varpi
t},\qquad\rho_x=2x\sgm'(\phi_\varpi+\varpi\phi_{\varpi\varpi}),
\eqno(3.19)$$
$$\rho_y=2y\sgm'(\phi_\varpi+\varpi\phi_{\varpi\varpi}),\qquad
\rho_z=\sgm^2\phi_{\varpi\varpi}.\eqno(3.20)$$ So (1.2) becomes
$$\phi_{\varpi t}-2\al'\phi_\varpi
+(\gm-4\al'\varpi)\phi_{\varpi\varpi}=0.\eqno(3.21)$$ Modulo $T_4$
in (1.12), the above equation is equivalent to:
$$\phi_ t+2\al'\phi+(\gm-4\al'\varpi)\phi_\varpi=0.\eqno(3.22)$$
Set
$$\td\psi=e^{2\al}\psi,\qquad\td\phi=e^{2\al}\phi.\eqno(3.23)$$
Then (3.15) and (3.22) are equivalent to the equations:
$$\td\psi_t+(\gm-4\al'\varpi)\td\psi_\varpi=0,\qquad
\td\phi_t+(\gm-4\al'\varpi)\td\phi_\varpi=0,\eqno(3.24)$$
respectively. So we have the solution
$$\td\phi=\Im(\td\psi)\lra \phi=e^{-2\al}\Im\left(e^{2\al}\sqrt{{\al'}'+(\al')^2+4^{-1}
-2\varpi}\right)\eqno(3.25)$$ for some one-variable function
$\Im$. Thus we have:\psp

{\bf Theorem 3.1}. {\it Let $\al$ be any function of $t$ and let
$\Im$ be arbitrary one-variable function.
 The following is a solution of the
equations (1.1)-(1.4) of  dynamic convection in a sea:
$$u=\al'
x-\frac{y}{2}+y\sqrt{{\al'}'+(\al')^2+\frac{1}{4}-\frac{2z}{x^2+y^2}},
\eqno(3.26)$$
$$v=\al'y+\frac{x}{2}-x\sqrt{{\al'}'+(\al')^2+\frac{1}{4}-\frac{2z}{x^2+y^2}},
\eqno(3.27)$$
$$w=\left(2(\al')^3+3\al'{\al'}'+\frac{{{\al'}'}'+\al'}{2}\right)(x^2+y^2)
-2\al' z,\eqno(3.28)$$
$$p=e^{-2\al}\Im\left(e^{2\al}\sqrt{{\al'}'+(\al')^2
+\frac{1}{4}-\frac{2z}{x^2+y^2}}\right),\eqno(3.29)$$
$$\rho
=-\frac{\Im'\left(e^{2\al}\sqrt{{\al'}'+(\al')^2
+\frac{1}{4}-\frac{2z}{x^2+y^2}}\right)}{(x^2+y^2)\sqrt{\al'+\al^2+\frac{1}{4}-\frac{2z}{x^2+y^2}}}
.\eqno(3.30)$$ }\pse

\section{Dimensional Reduction}

Suppose that $u,v,\zeta$ and $\eta$ are functions in $t,x,y$.
Assume
$$w=\zeta-(u_x+v_y)z,\qquad p=z+\eta,\qquad \rho=1.\eqno(4.1)$$
 Then the equations (1.1)-(1.4) are equivalent to the
following two-dimensional problem:
$$u_t+uu_x+vu_y+v=-\eta_x,\eqno(4.2)$$
$$v_t+uv_x+vv_y-u=-\eta_y.\eqno(4.3)$$
The compatibility $\eta_{xy}=\eta_{yx}$ gives
$$(u_y-v_x)_t+u(u_y-v_x)_x+v(u_y-v_x)_y
+(u_x+v_y)(u_y-v_x+1)=0.\eqno(4.4)$$

Suppose that $\vt$ is a function in $t,x,y$ such that
$$\vt_{xx}+\vt_{yy}=0\eqno(4.5)$$
(so $\vt$ is a time-dependent harmonic function). We assume
$$u=\vt_{xx},\qquad v=\vt_{xy}.\eqno(4.6)$$
Then (4.4) naturally holds. Indeed,
$$u_t+uu_x+vu_y+v=
\left(\vt_{xt}+2^{-1}(\vt_{xx}^2+\vt_{xy}^2)+\vt_y\right)_x,\eqno(4.7)$$
$$v_t+uv_x+vv_y-u=\left(\vt_{xt}+2^{-1}(\vt_{xx}^2+\vt_{xy}^2)+\vt_y\right)_y.
\eqno(4.8)$$ By (4.2) and (4.3), we take
$$\eta=-\vt_{xt}-\vt_y-\frac{1}{2}(\vt_{xx}^2+\vt_{xy}^2).\eqno(4.8)$$
Hence we have the following easy result:\psp

{\bf Proposition 4.1}. {\it Let $\vt$ and $\zeta$ be functions in
$t,x,y$ such that (4.5) holds. The following is a solution of the
equations (1.1)-(1.4) of  dynamic convection in a sea:
$$u=\vt_{xx},\qquad v=\vt_{xy},\qquad w=\zeta,\eqno(4.9)$$
$$\rho=1,\qquad p=z-\vt_{xt}-\vt_y-\frac{1}{2}(\vt_{xx}^2+\vt_{xy}^2).
\eqno(4.10)$$}

The above approach is the well-known rotation-free approach. We
are more interested in the approaches that the rotation may not be
zero. Let $f$ and $g$ be functions in $t,x,y$ that are linear in
$x,y$. Denote
$$\varpi=x^2+y^2.\eqno(4.11)$$
Consider
$$u=f+y\phi(t,\varpi),\qquad v=g
-x\phi(t,\varpi),\eqno(4.12)$$ where $\phi$ is a two-variable
function to be determined. Then
$$u_x=f_x+2xy\phi_\varpi,\qquad
u_y=f_y+\phi+2y^2\phi_\varpi,\eqno(4.13)$$
$$v_x=g_x-\phi-2x^2\phi_\varpi,\qquad
u_y=g_y-2xy\phi_\varpi.\eqno(4.14)$$ Thus
$$u_x+v_y=f_x+g_y, \qquad
u_y-v_x=f_y-g_x+2(\varpi\phi)_\varpi.\eqno(4.15)$$ For simplicity,
we assume
$$f=-\frac{\al'x}{2\al}-\frac{y}{2},\qquad g=\frac{x}{2}
x-\frac{\al'y}{2\al}\eqno(4.16)$$ for some functions $\al$ and
$\be$ of $t$. Then (4.4) becomes
$$(\varpi\phi)_{\varpi t}-\frac{\al'}{\al}
\varpi(\varpi\phi)_{\varpi\varpi}-\frac{\al'}{\al}
(\varpi\phi)_\varpi=0.\eqno(4.17)$$
 Hence
$$\phi=\frac{\gm+\Im(\al\varpi)}{\varpi}\eqno(4.18)$$
for some function $\gm$ of $t$ and one-variable function $\Im$.

Now (4.12), (4.16) and (4.18) imply
$$u=-\frac{\al'x}{2\al}-\frac{y}{2}
+\frac{(\gm+\Im(\al\varpi))y}{\varpi},\eqno(4.19)$$
$$v=\frac{x}{2}-\frac{\al'y}{2\al}-\frac{(\gm+\Im(\al\varpi))x}{\varpi}.
\eqno(4.20)$$ Moreover, (4.2) and (4.3) yield
$$\left(\frac{(\al')^2-2\al{\al'}'}{4\al^2}+\frac{1}{4}
\right)x +\frac{\gm' y}{\varpi}-x\phi^2 =-\eta_x,\eqno(4.21)$$
$$\left(\frac{(\al')^2-2\al{\al'}'}{4\al^2}+\frac{1}{4}
\right)y-\frac{\gm' x}{\varpi}-y\phi^2 =-\eta_y.\eqno(4.22)$$ Thus
$$\eta=\frac{1}{2}\int
\frac{(\gm+\Im(\al\varpi))^2d\varpi}{\varpi^2}
-\frac{1}{2}\left(\frac{(\al')^2-2\al{\al'}'}{4\al^2}+\frac{1}{4}
\right)\varpi+\gm' \arctan\frac{y}{x}.\eqno(4.23)$$\pse

{\bf Theorem 4.2}. {\it Let $\al,\gm$ be any functions of $t$.
Suppose that $\Im$ is an arbitrary one-variable function and
$\zeta$ is any function in $t,x,y$. The following is a solution of
the equations (1.1)-(1.4) of  dynamic convection in a sea:
$$u=-\frac{\al'x}{2\al}-\frac{y}{2}
+\frac{(\gm+\Im((x^2+y^2)\al))y}{x^2+y^2},\eqno(4.24)$$
$$v=\frac{x}{2}-\frac{\al'y}{2\al}-
\frac{(\gm+\Im((x^2+y^2)\al))x}{x^2+y^2},\eqno(4.25)$$
$$w=\frac{\al'}{\al}z+\zeta,\qquad
\rho=1,\eqno(4.26)$$
$$p=z+\frac{1}{2}\int
\frac{(\gm+\Im(\al\varpi))^2d\varpi}{\varpi^2}
-\frac{1}{2}\left(\frac{(\al')^2-2\al{\al'}'}{4\al^2}+\frac{1}{4}
\right)(x^2+y^2)+\gm' \arctan\frac{y}{x}\eqno(4.27)$$ with
$\varpi=x^2+y^2$.} \psp

Next we assume
$$u=\ves(t,x),\qquad v=\phi(t,x)+\psi(t,x)y,\eqno(4.28)$$
where $\ves,\;\phi$ and $\psi$ are functions in $t,x$ to be
determined. Substituting (4.28) into (4.4), we get
$$\phi_{tx}+\psi_{tx}y+\ves(\phi_{xx}+\psi_{xx}y)+(\phi+\psi
y)\psi_x+(\ves_x+\psi)(\phi_x+\psi_xy-1)=0,\eqno(4.29)$$
equivalently,
$$(\phi_t+\ves\phi_x+\phi\psi-\ves)_x-\psi=0,
\eqno(4.30)$$
$$(\psi_t+\ves\psi_x+\psi^2)_x=0.\eqno(4.31)$$
For simplicity, we take
$$\psi=-\al',\eqno(4.32)$$
a function of $t$.

Denote
$$\phi=\hat\phi+x.\eqno(4.33)$$
Then (4.30) becomes
$$(\hat\phi_t+\ves\hat\phi_x-\al'\hat\phi)_x=0.\eqno(4.34)$$
To solve the above equation, we assume
$$\ves=\frac{\be}{\hat\phi_x}-\frac{\vt_t(t,x)}{\vt_x(t,x)}\eqno(4.35)$$
for some functions $\be$ of $t$, and $\vt$ of $t$ and $x$.  We
have the following solution of (4.34):
$$\hat\phi=e^\al\Im(\vt)\lra\phi=e^\al\Im(\vt)+x\lra
v=e^\al\Im(\vt)+x-\al'y \eqno(4.36)$$ for another one-variable
function $\Im$. Moreover,
$$\ves=\frac{\be e^{-\al}}{\vt_x\Im'(\vt)}-
\frac{\vt_t}{\vt_x}.\eqno(4.37)$$ Note
\begin{eqnarray*} & &u_t+uu_x+vu_y+v
=\frac{(\be e^{-\al})'}{\vt_x\Im'(\vt)}- \frac{\be
e^{-\al}(\vt_{xt}\Im'(\vt)+\vt_t\vt_x{\Im'}'(\vt))
}{(\vt_x\Im'(\vt))^2}-
\frac{\vt_{tt}\vt_x-\vt_t\vt_{xt}}{\vt_x^2}-\al'y\\
& &+\left(\frac{\be e^{-\al}}{\vt_x\Im'(\vt)}-
\frac{\vt_t}{\vt_x}\right)\left(\frac{\be
e^{-\al}}{\vt_x\Im'(\vt)}- \frac{\vt_t}{\vt_x}\right)_x+
e^\al\Im(\vt)+x,\hspace{5.3cm}(4.38)
\end{eqnarray*}
$$v_t+uv_x+vv_y-u =((\al')^2-{\al'}')y +\be-\al'x.\eqno(4.39)$$
By (4.2) and (4.3),
\begin{eqnarray*}\eta&=&\int\left(\frac{\be
e^{-\al}(\vt_{xt}\Im'(\vt)+\vt_t\vt_x{\Im'}'(\vt))
}{(\vt_x\Im'(\vt))^2}+\frac{\vt_{tt}\vt_x-\vt_t\vt_{xt}}{\vt_x^2}
-\frac{(\be e^{-\al})'}{\vt_x\Im'(\vt)}-e^\al\Im(\vt)\right)dx
\\ &  &+\al' xy-\be
y+\frac{({\al'}'-(\al')^2)y^2-x^2}{2}-\frac{1}{2}\left(\frac{\be
e^{-\al}}{\vt_x\Im'(\vt)}-
\frac{\vt_t}{\vt_x}\right)^2.\hspace{3.9cm}(4.40)\end{eqnarray*}

{\bf Theorem 4.3}. {\it Let $\al,\be$ be functions of $t$ and let
$\Im$ be a one-variable function. Suppose that $\vt$ and $\zeta$
are functions in $t,x,y$. The following is a solution of the
equations (1.1)-(1.4) of  dynamic convection in a sea:
$$u=\frac{\be e^{-\al}}{\vt_x\Im'(\vt)}-
\frac{\vt_t}{\vt_x},\qquad v=e^\al\Im(\vt)+x-\al'y,\eqno(4.41)$$
$$w=\left(\al'+\frac{\be e^{-\al}(\vt_{xx}\Im'(\vt)+\vt_x^2{\Im'}'(\vt)
)}{(\vt_x\Im'(\vt))^2}+\frac{\vt_{xt}\vt_x-\vt_t\vt_{xx}}{\vt_x^2}\right)z+
\zeta,\qquad \rho=1,\eqno(4.42)$$
\begin{eqnarray*}p&=&z+\int\left(\frac{\be
e^{-\al}(\vt_{xt}\Im'(\vt)+\vt_t\vt_x{\Im'}'(\vt))
}{(\vt_x\Im'(\vt))^2}+\frac{\vt_{tt}\vt_x-\vt_t\vt_{xt}}{\vt_x^2}
-\frac{(\be e^{-\al})'}{\vt_x\Im'(\vt)}-e^\al\Im(\vt)\right)dx
\\ &  &+\al' xy-\be
y+\frac{({\al'}'-(\al')^2)y^2-x^2}{2}-\frac{1}{2}\left(\frac{\be
e^{-\al}}{\vt_x\Im'(\vt)}-
\frac{\vt_t}{\vt_x}\right)^2.\hspace{3.9cm}(4.43)\end{eqnarray*} }
\psp

Finally, we suppose that $\al,\be$ are functions of $t$ and $f,g$
are functions of $t,x,y$ that are linear homogeneous in $x$ and
$y$. Denote $\varpi=\al x+\be y$. Assume
$$u=f+\be \phi(t,\varpi),\qquad
v=g-\al\phi(t,\varpi).\eqno(4.44)$$ Then
$$u_y-v_x=f_y-g_x+(\al^2+\be^2)\phi_\varpi,\qquad
u_x+v_y=f_x+g_y.\eqno(4.45)$$ Now (4.4) becomes
\begin{eqnarray*}\hspace{1.4cm}& &f_{yt}-g_{xt}+(\al^2+\be^2)'\phi_\varpi+
(\al^2+\be^2)(\phi_{\varpi
t}+(\al'x+\be'y+\al f+\be g)\phi_{\varpi\varpi})\\
&&+(f_x+g_y)(f_y-g_x+1+(\al^2+\be^2)\phi_\varpi)=0.
\hspace{5.6cm}(4.46)\end{eqnarray*} In order to solve the above
equation, we assume
$$g_x=\vf,\qquad f_y=\vf-1,\eqno(4.47)$$
$$\al'x+\be'y+\al f+\be g=0\eqno(4.48)$$ for some function $\vf$ of $t$. The
equation (4.48) is equivalent to:
$$\al'+\al f_x+\vf\be=0\lra
f_x=-\frac{\al'+\vf\be}{\al},\eqno(4.49)$$
$$\be'+\be g_y+\al(\vf-1)=0\lra
g_y=-\frac{\be'+\al(\vf-1)}{\be}.\eqno(4.50)$$

Now (4.46) becomes
$$\phi_{\varpi
t}-\left(\frac{\al'+\vf\be}{\al}
+\frac{\be'+\al(\vf-1)}{\be}-\frac{(\al^2+\be^2)'}{\al^2+\be^2}\right)
\phi_\varpi=0.\eqno(4.51)$$ Thus we have the following solution:
$$\phi=\frac{\al+\be}{\al^2+\be^2}e^{\int(\al\be^{-1}(\vf-1)+\al^{-1}\be\vf)dt}
\Im'(\varpi),\eqno(4.52)$$ where $\Im$ is an arbitrary
one-variable function. Note
\begin{eqnarray*}
u_t+uu_x+vu_y+v&=&\frac{\al(\al+\be)}{\al^2+\be^2}
\left(\frac{\al\be'-\al'\be}{\al^2+b^2}-1\right)e^{\int(\al\be^{-1}(\vf-1)+\al^{-1}\be\vf)dt}
\Im'(\varpi)
\\& &+\left(\frac{2(\al')^2+(\vf\be)^2+3\al'\be\vf-\al(\vf\be)'
-\al{\al'}'}{\al^2}+\vf^2\right)x\\ & & + \left(\vf'-
\frac{(\vf-1)(\al'+\vf\be)}{\al}
-\frac{\vf(\be'+\al(\vf-1))}{\be}\right)y,\hspace{2.3cm}(4.53)
\end{eqnarray*}
\begin{eqnarray*}v_t+uv_x+vv_y-u&=&\frac{\be(\al+\be)}{\al^2+\be^2}
\left(\frac{\al\be'-\al'\be}{\al^2+b^2}-1\right)e^{\int(\al\be^{-1}(\vf-1)+\al^{-1}\be\vf)dt}
\Im'(\varpi)+[(\vf-1)^2\\ &&-\frac{\be((\vf-1)\al)'
+\be{\be'}'-2(\be')^2-((\vf-1)\al)^2-3\al\be'(\vf-1)}{\be^2}]y\\
& &+ \left(\vf'- \frac{(\vf-1)(\al'+\vf\be)}{\al}
-\frac{\vf(\be'+\al(\vf-1))}{\be}\right)x.\hspace{2.5cm}(4.54)
\end{eqnarray*}
By (4.2) and (4.3),
\begin{eqnarray*}\eta&=&\frac{y^2}{2}\left(\frac{\be((\vf-1)\al)'
+\be{\be'}'-2(\be')^2-((\vf-1)\al)^2-3\al\be'(\vf-1)}{\be^2}-(\vf-1)^2
\right)\\ & &-
\frac{x^2}{2}\left(\frac{2(\al')^2+(\vf\be)^2+3\al'\be\vf-\al(\vf\be)'
-\al{\al'}'}{\al^2}+\vf^2\right)+ [
\frac{(\vf-1)(\al'+\vf\be)}{\al}-\vf'\\ & &
+\frac{\vf(\be'+\al(\vf-1))}{\be}]xy +\frac{\al+\be}{\al^2+\be^2}
\left(1-\frac{\al\be'-\al'\be}{\al^2+b^2}\right)e^{\int(\al\be^{-1}(\vf-1)+\al^{-1}\be\vf)dt}
\Im(\varpi).\hspace{0.4cm}(4.55)
\end{eqnarray*}\pse

{\bf Theorem 4.4}. {\it Let $\al,\be,\vf$ be functions of $t$ and
let $\Im$ be a one-variable function. Suppose that $\zeta$ is
functions in $t,x,y$. The following is a solution of the equations
(1.1)-(1.4) of  dynamic convection in a sea:
$$u=(\vf-1)y-\frac{(\al'+\vf\be)x}{\al}+
\frac{\be(\al+\be)}{\al^2+\be^2}
e^{\int(\al\be^{-1}(\vf-1)+\al^{-1}\be\vf)dt} \Im'(\al x+\be
y),\eqno(4.56)$$
$$v=\vf x-\frac{(\be'+(\vf-1)\al)y}{\be}-
\frac{\al(\al+\be)}{\al^2+\be^2}e^{\int(\al\be^{-1}(\vf-1)+\al^{-1}\be\vf)dt}
\Im'(\al x+\be y),\eqno(4.57)$$
$$w=\left(\frac{\al'+\vf\be}{\al}
+\frac{\be'+(\vf-1)\al}{\be}\right)z+\zeta,\qquad\rho=1,\eqno(4.58)$$
\begin{eqnarray*}& &p=z+\frac{y^2}{2}\left(\frac{\be((\vf-1)\al)'
+\be{\be'}'-2(\be')^2-((\vf-1)\al)^2-3\al\be'(\vf-1)}{\be^2}-(\vf-1)^2
\right)\\ & &-
\frac{x^2}{2}\left(\frac{2(\al')^2+(\vf\be)^2+3\al'\be\vf-\al(\vf\be)'
-\al{\al'}'}{\al^2}+\vf^2\right)+xy [
\frac{(\vf-1)(\al'+\vf\be)}{\al} -\vf'\\
& &+\frac{\vf(\be'+\al(\vf-1))}{\be}] +\frac{\al+\be}{\al^2+\be^2}
\left(1-\frac{\al\be'-\al'\be}{\al^2+b^2}\right)e^{\int(\al\be^{-1}(\vf-1)+\al^{-1}\be\vf)dt}
\Im(\al x+\be y).\hspace{0.4cm}(4.59)
\end{eqnarray*}}

\bibliographystyle{amsplain}

\end{document}